\documentclass[aps,twocolumn,superscriptaddress,amsmath,longbibliography]{revtex4-1}
\usepackage{graphicx,color,hyperref}
\usepackage{soul}
\usepackage{amsfonts}
\usepackage{amssymb}
\usepackage{amsmath}
\usepackage{amsthm}
\usepackage{bm}
\usepackage{braket}
\usepackage{enumerate}

\begin{document}

\title{Spin-mediated Mott excitons}

\author{T.-S. Huang}
\affiliation{Joint Quantum Institute, University of Maryland, College Park, MD 20742, USA}
\affiliation{Condensed Matter Theory Center, University of Maryland, College Park, MD 20742, USA}

\author{C. L. Baldwin}
\affiliation{Joint Quantum Institute, University of Maryland, College Park, MD 20742, USA}
\affiliation{National Institute of Standards and Technology, Gaithersburg, MD 20899, USA}

\author{M. Hafezi}
\affiliation{Joint Quantum Institute, University of Maryland, College Park, MD 20742, USA}

\author{V. Galitski}
\affiliation{Joint Quantum Institute, University of Maryland, College Park, MD 20742, USA}
\affiliation{Condensed Matter Theory Center, University of Maryland, College Park, MD 20742, USA}

\date{\today}

\begin{abstract}

Motivated by recent experiments on Mott insulators, in both iridates and ultracold atoms, we theoretically study the effects of magnetic order on the Mott-Hubbard excitons. In particular, we focus on spin-mediated doublon-holon pairing in Hubbard materials. We use several complementary theoretical techniques: mean-field theory to describe the spin degrees of freedom, the self-consistent Born approximation to characterize individual charge excitations across the Hubbard gap, and the Bethe-Salpeter equation to identify bound states of doublons and holons. The binding energy of the Mott exciton is found to increase with increasing the N{\'e}el order parameter, while the exciton mass decreases. We observe that these trends rely significantly on the retardation of the effective interaction, and require consideration of multiple effects from changing the magnetic order. Our results are consistent with the key qualitative trends observed in recent experiments on iridates. Moreover, the findings could have direct implications on ultracold atom Mott insulators, where the Hubbard model is the exact description of the system and the microscopic degrees of freedom can be directly accessed.

\end{abstract}

\maketitle

\section{Introduction} \label{sec:introduction}

The physics of excitons in semiconductors, i.e., bound states of electrons and holes, is by now well-established~\cite{Haug1984Electron,Haug2004}.
Excitons play an essential role in technologies such as light-emitting diodes~\cite{Burroughes1990Light}, organic solar cells~\cite{Gregg2003Excitonic}, and photodetectors~\cite{Itkis2006Bolometric}, among others.
Furthermore, there has recently been tremendous interest in hybridizing exciton states with photon modes in optical cavities~\cite{Carusotto2013Quantum}.
Such exciton polaritons can form (non-equilibrium) Bose-Einstein condensates at remarkably high temperatures, even room temperature~\cite{Kasprzak2006Bose,Keeling2007Collective,Deng2010Exciton,Byrnes2014Exciton,Plumhof2014Room}.

Given these applications, it is important to study the properties of excitons in systems other than conventional semiconductors.
It has been convincingly established that excitonic states do exist in strongly correlated materials such as Mott insulators~\cite{Gallagher1997Excitons,Essler2001Excitons,Wrobel2002Excitons,Matsueda2005Excitonic,Ono2005Direct,Gossling2008Mott,AlHassanieh2008Excitons,Novelli2012Ultrafast,Gretarsson2013Crystal,Kim2014Excitonic}, yet essential aspects of Mott excitons remain poorly understood.

\begin{figure}[t]
\centering
\includegraphics[width=1.0\columnwidth]{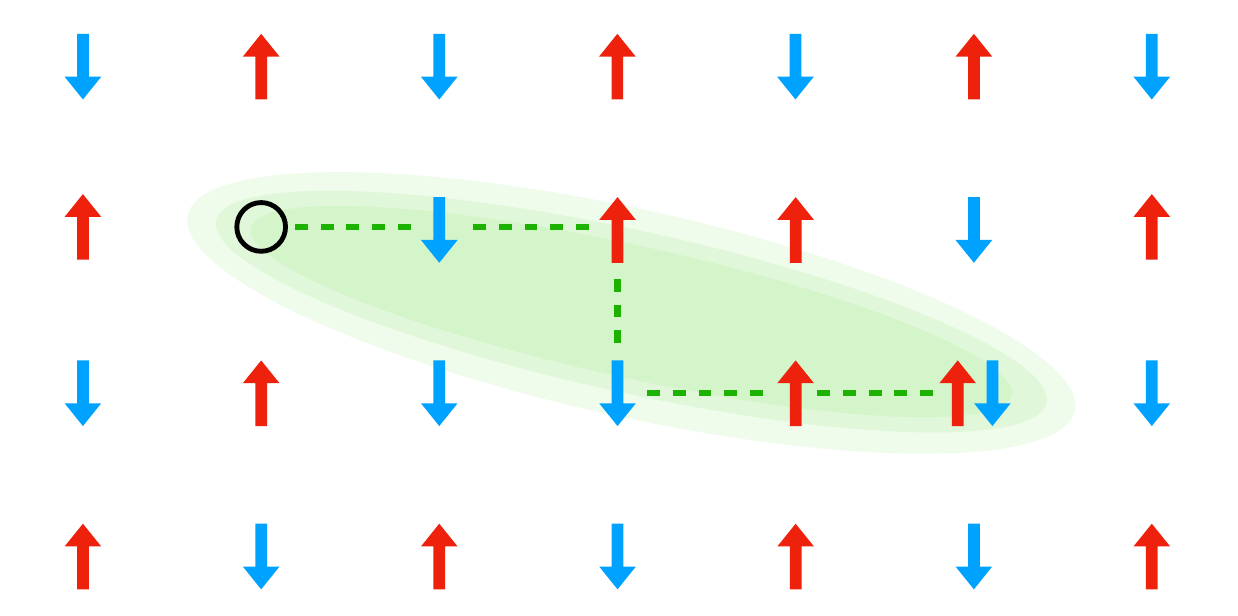}
\caption{Sketch of the physics underlying spin-mediated exciton formation in a Mott insulator. A string of flipped spins (green line) connects the hole and double-occupancy. The energy cost associated to the string binds them together.}
\label{fig:exciton_cartoon}
\end{figure}

For example, it is known that Mott insulators are often antiferromagnetic at low temperature, but very little work has been done to understand how and to what extent the presence of such order affects exciton properties.
The qualitative role of magnetization is sketched in Fig.~\ref{fig:exciton_cartoon} -- charges remain bound so as to minimize the number of spins disrupted by their motion -- but a quantitative description has been lacking. 

Recent experiments have begun to investigate this question.
In Refs.~\cite{Alpichshev2015Confinement,Alpichshev2017Origin}, pump-probe experiments were performed on the Mott insulator Na\textsubscript{2}IrO\textsubscript{3} both with and without magnetic order (controlled by varying temperature or applying an intermediate pulse).
The authors concluded that the binding energy and exciton mass are both enhanced by the presence of magnetization.
Ref.~\cite{Terashige2019Doublon} similarly observed that the binding energy increases with the spin-spin interaction strength in cuprates.
See also Ref.~\cite{Eckstein2016Ultra}, which found that the relaxation time in Mott insulators decreases with increasing spin correlations.

The same question can apply to Mott insulators in synthetic quantum systems, such as ultracold gases.
By loading fermionic atoms into an optical lattice and tuning their interactions, the Fermi-Hubbard model can be synthesized experimentally~\cite{Schneider2008Metallic,Bloch2008Many,Esslinger2010Fermi,Bloch2012Quantum,Gross2017Quantum}.
Unlike condensed matter systems, such as the iridates, neutral fermionic atoms in an optical lattice are genuinely described by the Hubbard Hamiltonian, without any additional effects arising from longer-range Coulomb interactions, phonons, etc.
Researchers have quite recently begun investigating the interplay of spin and charge degrees of freedom in this setting~\cite{Parsons2016Site,Boll2016Spin,Cheuk2016Observation,Hilker2017Revealing,Salomon2019Direct,Koepsell2019Imaging,Vijayan2020Time} (note that here the ``charge'' excitations are not actually charged).

In this paper, we perform a theoretical study of the role of magnetic order in Mott excitons, the first such to our knowledge. As depicted in Fig.~\ref{fig:exciton_cartoon}, in an anti-ferromagnetic background, a hole and doubly occupied site can bind through a string of flipped spins. Such Mott excitons differ from conventional excitons formed by Coulomb interaction in two aspects.
First, the spin-mediated interaction is far from instantaneous, and second, the individual charges are themselves renormalized by spin fluctuations. We shall demonstrate that both effects are necessary ingredients in the trends reported here.

Given the complexity of the problem, our analysis requires multiple stages. We first use slave particles to isolate spin and charge degrees of freedom, then describe the spin dynamics by mean-field theory, calculate the dispersion of charges self-consistently, and finally characterize excitonic states via the Bethe-Salpeter equation.
Many of the steps in this program are analogous to those in Ref.~\cite{Han2016Charge}, which studied charge dynamics in the Hubbard model.
The good agreement between the results of Ref.~\cite{Han2016Charge} and alternate numerical methods lends support to the present approach.

Our key finding is that larger magnetization leads to an increased binding energy of the Hubbard exciton but a decreased mass.
This observation is in some tension with interpretations of recent experiments~\cite{Alpichshev2017Origin}.
It also stands in contrast to conventional Coulomb-mediated excitons, where the binding energy and mass are proportional to each other.

Note that the formation of Mott excitons is closely related to the Cooper pairing of holes in high-$T_c$ superconductors.
Similar treatments of hole-hole binding can be found in the corresponding literature~\cite{Shraiman1989Mean,Kuchiev1993Large,Belinicher1995Hole,Belinicher1997Two,Brugger2006Magnon}.
Nonetheless, there are differences between the two problems, as we discuss below.
Moreover, to the best of our knowledge, a study of how the bound state properties change as a function of magnetization has not yet been carried out.

In the following Sec.~\ref{sec:formalism}, we describe the steps of our analysis in detail. Results are presented in Sec.~\ref{sec:results}, and Sec.~\ref{sec:conclusion} concludes.

\section{Formalism \& methods} \label{sec:formalism}

Our starting point is the 2D Fermi-Hubbard model, which by now needs no introduction:
\begin{equation} \label{eq:Hubbard_model_def}
H_{\textrm{Hub}} = -t \sum_{\langle ij \rangle , \sigma} c_{i \sigma}^{\dag} c_{j \sigma} + U \sum_i n_{i \uparrow} n_{i \downarrow},
\end{equation}
where $\sigma \in \{ \uparrow , \downarrow \}$ and $\langle ij \rangle$ denotes nearest-neighbor sites on a square lattice.
$c_{i \sigma}$ is the usual electron annihilation operator and $n_{i \sigma} \equiv c_{i \sigma}^{\dag} c_{i \sigma}$.
We shall consider the system at half filling in the $U \gg t$ limit.

It is well-known that in this limit, the Hubbard model features two types of excitations, associated with the transport of charge and spin respectively~\cite{Hirsch1985Two,Kotliar1986New,White1989Numerical,Lee1989Gauge}.
Furthermore, the charge excitations can be either positive or negative, corresponding to sites with zero or two electrons, and their creation comes with a large energy cost of order $U$.
By analogy with conventional semiconductors, we thus expect this system to support well-defined excitons in the dilute-charge limit.
However, long-wavelength spin excitations do not come with an energy cost, and their presence plays a significant role in determining the exciton properties.

There are many formalisms with which to study the Hubbard model~\cite{Ogawa1975Gutzwiller,MacDonald1988Expansion,Li1989Spin,Lee2006Doping}.
Since our focus is on the motion of only a few charges within a background of spin excitations, the slave-particle formalism is particularly well-suited~\cite{SchmittRink1988Spectral,Kane1989Motion,Han2016Charge}.
The steps of our calculation are as follows:
\begin{enumerate}[i)]
    \item Express the Hamiltonian in terms of slave particles -- doublons, holons, \& spinons -- and reduce to the t-J model following the standard procedure~\cite{Auerbach1994}.
    \item Treat the Heisenberg interaction within semiclassical and mean-field approximations, while neglecting the back action of doublons and holons on the magnetic order.
    \item Calculate the dispersion of individual doublons and holons in the magnetic background via the self-consistent Born approximation.
    \item Calculate exciton properties using the Bethe-Salpeter equation.
\end{enumerate}

The major limitation of this program is our approximate description of the magnetic order.
Thus we do not claim to have quantitatively accurate results, especially at small magnetization.
That said, we do expect that the qualitative trends seen here are accurate, including near the equilibrium value of magnetization, for which mean field theory is known to work reasonably well (see Ref.~\cite{Han2016Charge} and references therein).

\subsection{Slave particles}

\begin{figure}[t]
\centering
\includegraphics[width=1.0\columnwidth]{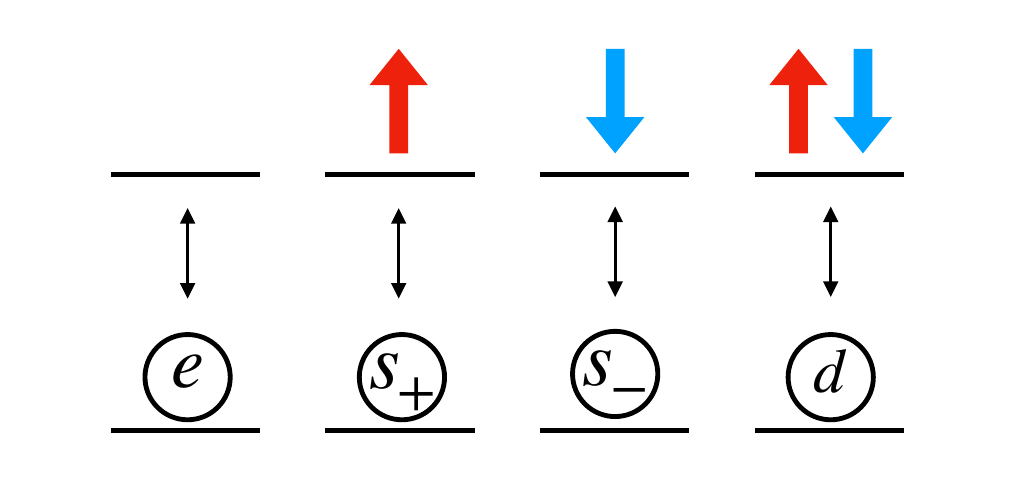}
\caption{Sketch of the slave-particle formalism. Each of the four possible occupancies on a site corresponds to a different type of fictitious particle.}
\label{fig:slave_particle_cartoon}
\end{figure}

In the slave-particle formalism, we express the electron operator as (with $\sigma = \pm 1$)
\begin{equation} \label{eq:slave_particle_def}
c_{i \sigma} = s_{i,-\sigma}^{\dag} d_i + \sigma e_i^{\dag} s_{i \sigma},
\end{equation}
where $d_i$ and $e_i$ are fermionic operators and $s_{i \sigma}$ is bosonic.
One can confirm that Eq.~\eqref{eq:slave_particle_def} is consistent with the commutation relations.
A site with a $d$ particle is to be interpreted as a site with two electrons (a ``doublon''), a site with an $e$ particle is to be interpreted as an empty site (a ``holon''), and a site with an $s_{\sigma}$ particle is one with a single electron having spin $\sigma$ (a ``spinon'').
See Fig.~\ref{fig:slave_particle_cartoon}.
The physical content of Eq.~\eqref{eq:slave_particle_def} is then clear: removing an electron of given spin is equivalent to replacing the doublon with the opposite spinon if the site is doubly-occupied and replacing the spinon with a holon if the site is singly-occupied (otherwise the state is annihilated).
Note that since every site is in one of the four states -- empty, spin-up, spin-down, doubly-occupied -- there must be exactly one of the fictitious particles on each site:
\begin{equation} \label{eq:slave_particle_constraint}
d_i^{\dag} d_i + e_i^{\dag} e_i + s_{i \uparrow}^{\dag} s_{i \uparrow} + s_{i \downarrow}^{\dag} s_{i \downarrow} = 1, \; \forall i.
\end{equation}
The original Hamiltonian clearly preserves this relationship.

Substituting Eq.~\eqref{eq:slave_particle_def} into Eq.~\eqref{eq:Hubbard_model_def}, we have that
\begin{equation} \label{eq:slave_particle_Hamiltonian}
\begin{aligned}
H_{\textrm{Hub}} =& -t \sum_{\langle ij \rangle , \sigma} \big( d_i^{\dag} d_j - e_i^{\dag} e_j \big) s_{j \sigma}^{\dag} s_{i \sigma} + U \sum_i d_i^{\dag} d_i \\
& -t \sum_{\langle ij \rangle , \sigma} \sigma \big( d_i^{\dag} e_j^{\dag} s_{i, -\sigma} s_{j \sigma} + e_i d_j s_{i \sigma}^{\dag} s_{j, -\sigma}^{\dag} \big) .
\end{aligned}
\end{equation}
Note that the first line preserves the number of doublons and holons, whereas the second line does not.

At large $U$, the second line of Eq.~\eqref{eq:slave_particle_Hamiltonian} can be treated by perturbation theory in $t/U$.
The method as applied here is standard, and can be found in, e.g., Ref.~\cite{Auerbach1994}.
We obtain the t-J model:
\begin{equation} \label{eq:t_J_model_def}
\begin{aligned}
H_{\textrm{tJ}} =& -t \sum_{\langle ij \rangle , \sigma} \big( d_i^{\dag} d_j - e_i^{\dag} e_j \big) s_{j \sigma}^{\dag} s_{i \sigma} + U \sum_i d_i^{\dag} d_i \\
& -J \sum_{\langle ij \rangle } \big( s_{i \uparrow}^{\dag} s_{j \downarrow}^{\dag} - s_{i \downarrow}^{\dag} s_{j \uparrow}^{\dag} \big) \big( s_{j \downarrow} s_{i \uparrow} - s_{j \uparrow} s_{i \downarrow} \big) ,
\end{aligned}
\end{equation}
where $J \equiv 4t^2 / U$.
Strictly speaking, Eq.~\eqref{eq:t_J_model_def} should include additional next-nearest-neighbor terms, as well as a direct interaction between nearest-neighbor doublons and holons, but these are commonly neglected.

\subsection{Magnetic ordering}

The second line of Eq.~\eqref{eq:t_J_model_def} is precisely the antiferromagnetic Heisenberg Hamiltonian, expressed in terms of Schwinger bosons (here the spinons $s_{i \sigma}$)~\cite{Auerbach1994}.
We are specifically considering N\'eel-ordered ground states, with positive magnetization on the $A$ sublattice and negative on $B$.
It is thus convenient to express $s_{i \uparrow}$ ($s_{i \downarrow}$) in terms of $s_{i \downarrow}$ ($s_{i \uparrow}$) for $i \in A$ ($B$), using Eq.~\eqref{eq:slave_particle_constraint} --- this is equivalent to using \textit{Holstein-Primakoff} rather than Schwinger bosons~\cite{Auerbach1994}:
\begin{equation} \label{eq:Holstein_Primakoff_relationship}
\begin{aligned}
s_{i \uparrow} &= \left[ 1 - s_{i \downarrow}^{\dag} s_{i \downarrow} \right]^{1/2}, \quad i \in A, \\
s_{i \downarrow} &= \left[ 1 - s_{i \uparrow}^{\dag} s_{i \uparrow} \right]^{1/2}, \quad i \in B.
\end{aligned}
\end{equation}
We neglect $d_i^{\dag} d_i$ and $e_i^{\dag} e_i$ in using Eq.~\eqref{eq:slave_particle_constraint} because we are interested in the dilute-charge limit.
From here on, we shall simply write $s_i$ in place of $s_{i \downarrow}$ ($s_{i \uparrow}$) for $i \in A$ ($B$).
On both sublattices, the $s_i$ boson represents a fluctuation relative to perfect N\'eel order.

Note that Eq.~\eqref{eq:Holstein_Primakoff_relationship} cannot be an exact equality because it does not respect the fact that $[s_{i \sigma}, s_{i \sigma}^{\dag}] = 1$.
It is more of a semiclassical approximation, valid in the limit of large spin $S$.
Of course, the case $S = 1/2$ under consideration here is far from large, but all other analytical techniques of which we are aware for treating long-range order via slave particles (such as Bose-Einstein condensation of the Schwinger bosons~\cite{Sarker1989Bosonic,Sarker1990Mean}) have the same regime of validity.
We refer for Ref.~\cite{Auerbach1994} for more details.

Inserting Eq.~\eqref{eq:Holstein_Primakoff_relationship} into $H_{\textrm{tJ}}$  does not yield a solvable Hamiltonian on its own.
Thus to progress further, we make a mean-field approximation by expanding $H_{\textrm{tJ}}$ to first order in $s_i^{\dag} s_i - \langle s_i^{\dag} s_i \rangle$ (this is also reasonable in the semiclassical limit).
With $\langle s_i^{\dag} s_i \rangle = 1/2 - m$, where $m$ is the N\'eel magnetization, we find that
\begin{equation} \label{eq:mean_field_Hamiltonian}
\begin{aligned}
H_{\textrm{tJ}} \approx & -\sqrt{\frac{1 + 2m}{2}} t \sum_{\langle ij \rangle} \big( d_i^{\dag} d_j - e_i^{\dag} e_j \big) \big( s_i + s_j^{\dag} \big) \\
& + 4 (1 + 2m) J \sum_i s_i^{\dag} s_i + \frac{1 + 2m}{2} J \sum_{\langle ij \rangle} \big( s_i^{\dag} s_j^{\dag} + s_i s_j \big) \\
& + U \sum_i d_i^{\dag} d_i - NJ (1 + 2m)^2.
\end{aligned}
\end{equation}
The second line is diagonalized by passing to momentum space and performing a Bogoliubov transformation, leading to the final Hamiltonian (neglecting constant terms)
\begin{equation} \label{eq:working_Hamiltonian}
\begin{aligned}
H_{\textrm{tJ}} \approx & -\sqrt{\frac{1 + 2m}{2N}} t \sum_{kq} d_{k+q}^{\dag} d_k \big( M_{kq} \beta_q + M_{k+q,-q} \beta_{-q}^{\dag} \big) \\
& + \sqrt{\frac{1 + 2m}{2N}} t \sum_{kq} e_{k+q}^{\dag} e_k \big( M_{kq} \beta_q + M_{k+q,-q} \beta_{-q}^{\dag} \big) \\
& + U \sum_q d_q^{\dag} d_q + \sum_q \omega_q \beta_q^{\dag} \beta_q,
\end{aligned}
\end{equation}
where the sum is over the 2D Brillouin zone ($N$ is the number of lattice sites) and $\beta_q \equiv u_q s_q + v_q s_{-q}^{\dag}$ is the transformed spinon operator.
With
\begin{equation} \label{eq:Bogoliubov_u_coefficient}
u_q \equiv \sqrt{\frac{1}{2} \left( 1 + \frac{1}{\sqrt{1 - \gamma_q^2}} \right) },
\end{equation}
\begin{equation} \label{eq:Bogoliubov_v_coefficient}
v_q \equiv \textrm{sgn}[\gamma_q] \sqrt{\frac{1}{2} \left( \frac{1}{\sqrt{1 - \gamma_q^2}} - 1 \right) },
\end{equation}
where
\begin{equation} \label{eq:hopping_coefficient}
\gamma_q \equiv \frac{1}{2} \big( \cos{q_x} + \cos{q_y} \big) ,
\end{equation}
the frequencies $\omega_q$ and vertices $M_{kq}$ entering into Eq.~\eqref{eq:working_Hamiltonian} are given by
\begin{equation} \label{eq:spinon_frequencies}
\omega_q = 4 (1 + 2m) J \sqrt{1 - \gamma_q^2},
\end{equation}
\begin{equation} \label{eq:spinon_charge_coupling}
M_{kq} = 4 \gamma_k u_q - 4 \gamma_{k+q} v_q.
\end{equation}

Normally one would determine $m$ self-consistently from the ground state spinon occupation: $1/2 - m = N^{-1} \sum_q \langle s_q^{\dag} s_q \rangle = N^{-1} \sum_q v_q^2$.
This is known to give $m \approx 0.3$ for a 2D square lattice~\cite{Han2016Charge}.
We shall instead treat $m$ as an independent parameter --- this is an approximate (albeit crude) means of estimating exciton properties as a function of magnetization.

\subsection{Self-consistent Born approximation}

\begin{figure}[t]
\centering
\includegraphics[width=1.0\columnwidth]{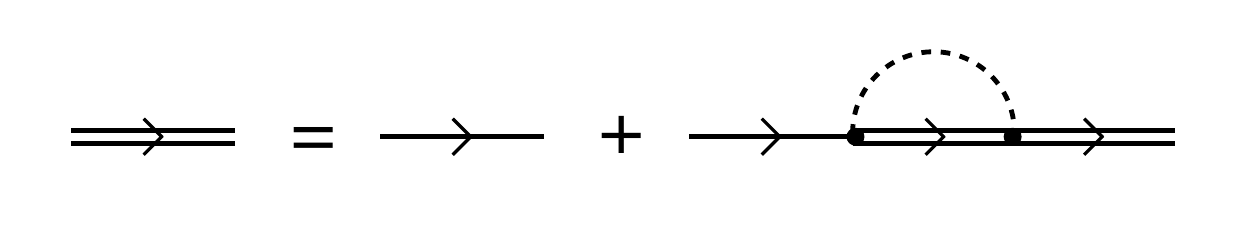}
\caption{The self-consistent Born approximation (SCBA) for the single-particle propagator (either doublon-doublon or holon-holon). The solid single line is the free propagator, in this case simply $G_k^0(\epsilon) = 1/\epsilon$, and the solid double line is the full propagator. The dashed line is the spinon propagator and the black dot is the vertex, corresponding to the Hamiltonian in Eq.~\eqref{eq:working_Hamiltonian}.}
\label{fig:SCBA}
\end{figure}

The self-consistent Born approximation (SCBA) gives the doublon-doublon and holon-holon propagators by the integral equation in Fig.~\ref{fig:SCBA}.
The same equation holds for each propagator separately.
This approximation is expected to be accurate in the dilute-charge limit, where the charge dynamics is strongly affected by spinons but not vice-versa.

In terms of the doublon/holon self-energy $\Sigma_k(\epsilon)$, Fig.~\ref{fig:SCBA} translates to (after a frequency integration)
\begin{equation} \label{eq:SCBA_self_energy}
\Sigma_k(\epsilon) = \frac{(1 + 2m) t^2}{2N} \sum_q \frac{M_{kq}^2}{\epsilon - \omega_q - \Sigma_{k-q}(\epsilon - \omega_q)}.
\end{equation}
The quasiparticle spectrum $\epsilon_k$ is given by the solution to $\Sigma_k(\epsilon_k) = \epsilon_k$.

Eq.~\eqref{eq:SCBA_self_energy} can be solved quite efficiently.
Note that all $\omega_q$ are positive~\footnote{More precisely, all $\omega_q$ are non-negative, but the vertex vanishes on those momenta such that $\omega_q = 0$.}, thus Eq.~\eqref{eq:SCBA_self_energy} in fact expresses $\Sigma_k(\epsilon)$ in terms of the self-energy at lower frequencies.
We start at sufficiently negative $\epsilon$, below which we approximate $\Sigma_k(\epsilon) \approx (1 + 2m)t^2/2N \sum_q M_{kq}^2 / (\epsilon - \omega_q)$, and then compute the self-energy at incrementally higher frequencies in terms of the previous values.
To help avoid numerical errors, we add a small imaginary part (namely $0.1i (1 + 2m) J$) to $\epsilon$.

\begin{figure}[t]
\centering
\includegraphics[width=0.7\columnwidth]{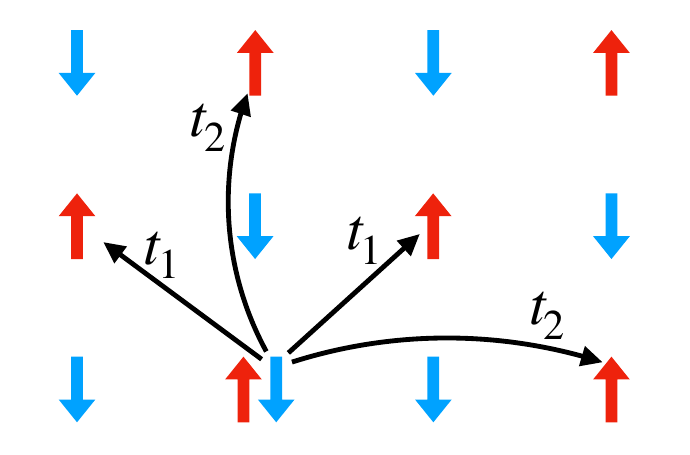}
\caption{Illustration of hopping parameters $t_1$ and $t_2$ for the approximation to the quasiparticle dispersion in Eq.~\eqref{eq:approximate_single_particle_dispersion}. Singly-occupied sites show the background magnetic order in which the double occupancy hops.}
\label{fig:effective_hopping}
\end{figure}

While one could proceed using the full $\Sigma_k(\epsilon)$, it has been found that the quasiparticle dispersion can be well-approximated by the form~\cite{Martinez1991Spin}
\begin{equation} \label{eq:approximate_single_particle_dispersion}
\begin{aligned}
\epsilon_k =& \, -2t_1 \big( \cos{(k_x + k_y)} + \cos{(k_x - k_y)} \big) \\
& \quad - 2t_2 \big( \cos{(2k_x)} + \cos{(2k_y)} + 2 \big) .
\end{aligned}
\end{equation}
This expression has a clear physical interpretation: $t_1$ is the amplitude for performing a two-step hop along the diagonals of the lattice, and $t_2$ is the amplitude for a two-step hop along the principal axes (see Fig.~\ref{fig:effective_hopping}).
Thus in what follows, we shall use for the single-particle propagators the simpler expression
\begin{equation} \label{eq:simple_single_particle_propagator}
G_k(\epsilon) = \frac{1}{\epsilon - (1 - i0) \epsilon_k},
\end{equation}
with $\epsilon_k$ given by Eq.~\eqref{eq:approximate_single_particle_dispersion}.

\begin{figure}[t]
\centering
\includegraphics[width=1.0\columnwidth]{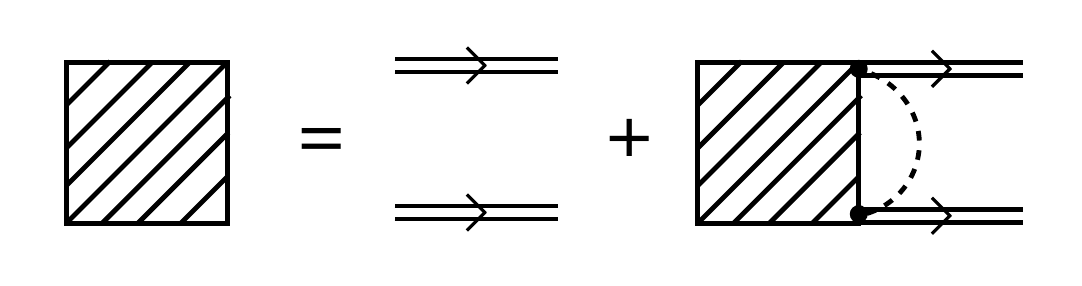}
\caption{The integral equation which determines the two-particle Green's function, within the ladder approximation. The hatched square is the Green's function, and all other symbols are as in Fig.~\ref{fig:SCBA}.}
\label{fig:Bethe_Salpeter}
\end{figure}

\subsection{Bethe-Salpeter equation}

We next consider the two-particle Green's function ($T$ denotes time ordering)
\begin{equation} \label{eq:two_body_GF_position}
\begin{aligned}
& \mathcal{G}_{j_d j_e; j_d' j_e'}(t_d, t_e; t_d', t_e') \\
& \qquad \quad \equiv - \big< T d_{j_d}(t_d) e_{j_e}(t_e) e_{j_e'}(t_e')^{\dag} d_{j_d'}(t_d')^{\dag} \big> ,
\end{aligned}
\end{equation}
and its Fourier transform $\mathcal{G}_{k_d k_e; k_d' k_e'}(\epsilon_d, \epsilon_e; \epsilon_d', \epsilon_e')$.
Due to translational invariance, $\mathcal{G}$ depends only on differences in position and time, which we choose to parametrize by the relative coordinates
\begin{equation} \label{eq:relative_coordinates_position}
\begin{aligned}
j \equiv & \; j_d - j_e, \\
j' \equiv & \; j_d' - j_e', \\
r \equiv & \; \frac{j_d + j_e}{2} - \frac{j_d' + j_e'}{2},
\end{aligned}
\end{equation}
with relative times defined analogously. 
The corresponding momenta are
\begin{equation} \label{eq:relative_coordinates_momentum}
\begin{aligned}
k =& \; \frac{k_d - k_e}{2}, \\
k' =& \; \frac{k_d' - k_e'}{2}, \\
K =& \; k_d + k_e = k_d' + k_e'.
\end{aligned}
\end{equation}
We will use absolute and relative momenta interchangeably, depending on notational convenience, with Eq.~\eqref{eq:relative_coordinates_momentum} always giving the relationship between the two.
$\mathcal{G}_{k_d k_e; k_d' k_e'}(\epsilon_d, \epsilon_e; \epsilon_d', \epsilon_e')$ will often be written as $\mathcal{G}_{k k' ; K}(\epsilon, \epsilon'; E)$.

Within the ladder approximation, $\mathcal{G}$ is determined by the integral equation of Fig.~\ref{fig:Bethe_Salpeter}.
Written out,
\begin{widetext}
\begin{equation} \label{eq:two_body_GF_equation}
\mathcal{G}_{k k'; K}(\epsilon, \epsilon'; E) = G_{k_d}(\epsilon_d) G_{k_e}(\epsilon_e) \Bigg[ \delta_{kk'} \, + \, \frac{(1 + 2m) t^2}{2N} \sum_q \int \frac{\textrm{d}\omega}{2\pi i} \left( \frac{M_{k_d - q, q} M_{k_e, q}}{\omega - (1 - i0) \omega_q} - \frac{M_{k_d, -q} M_{k_e + q, -q}}{\omega + (1 - i0) \omega_q} \right) \mathcal{G}_{k-q, k'; K}(\epsilon - \omega, \epsilon'; E) \Bigg],
\end{equation}
where $G_k(\epsilon)$ is given by Eq.~\eqref{eq:simple_single_particle_propagator} and the vertices $M_{kq}$ are as in Eq.~\eqref{eq:working_Hamiltonian}.

Since our goal is to identify bound states, we reduce Eq.~\eqref{eq:two_body_GF_equation} to the Bethe-Salpeter equation.
The details of this approach can be found in Ref.~\cite{Nakanishi1969General}.
We assume that $\mathcal{G}$ has an isolated pole in the total energy $E$, near which it has the form
\begin{equation} \label{eq:Bethe_Salpeter_form}
\mathcal{G}_{kk'; K}(\epsilon, \epsilon'; E) \sim -i \frac{\psi_k(\epsilon) \overline{\psi}_{k'}(\epsilon')}{E - (1 - i0) E_b},
\end{equation}
where the ``wavefunction'' $\psi_k(\epsilon)$, its time-reversed partner $\overline{\psi}_k(\epsilon)$, and the bound state energy $E_b$ remain to be determined.
Inserting this ansatz into both sides of Eq.~\eqref{eq:two_body_GF_equation} and equating the residues at $E_b$ on each side, we obtain a non-linear eigenvalue problem (the Bethe-Salpeter equation):
\begin{equation} \label{eq:Bethe_Salpeter_equation}
\psi_k(\epsilon) = G_{k_d} \left( \frac{E_b}{2} + \epsilon \right) G_{k_e} \left( \frac{E_b}{2} - \epsilon \right) \frac{(1 + 2m) t^2}{2N} \sum_q \int \frac{\textrm{d}\omega}{2\pi i} \left( \frac{M_{k_d - q, q} M_{k_e, q}}{\omega - (1 - i0) \omega_q} - \frac{M_{k_d, -q} M_{k_e + q, -q}}{\omega + (1 - i0) \omega_q} \right) \psi_{k-q}(\epsilon - \omega).
\end{equation}
$E_b$ and $\psi_k(\epsilon)$ are given by the solution to Eq.~\eqref{eq:Bethe_Salpeter_equation}.
Note that they will depend on the center-of-mass momentum $K$.

$\psi_k(\epsilon)$ is the bound state wavefunction in a quite literal sense: it is the Fourier transform of
\begin{equation} \label{eq:generic_bound_state_wavefunction}
\psi_j(t) = \langle 0 | T d_{j}(t) e_0(0) | b \rangle ,
\end{equation}
where $| b \rangle$ denotes the bound state and $| 0 \rangle$ denotes the ground state.
Note that $t = 0$ is of particular interest, since it gives the amplitude for \textit{simultaneously} observing the holon at site 0 and the doublon at site $j$.
Thus to simplify the problem, we integrate Eq.~\eqref{eq:Bethe_Salpeter_equation} over $\epsilon$, and furthermore, make the ansatz
\begin{equation} \label{eq:Bethe_Salpeter_frequency_ansatz}
\psi_k(\epsilon) = -G_{k_d} \left( \frac{E_b}{2} + \epsilon \right) G_{k_e} \left( \frac{E_b}{2} - \epsilon \right) \big( E_b - \epsilon_{k_d} - \epsilon_{k_e} \big) \Psi_k,
\end{equation}
with $\Psi_k$ independent of $\epsilon$.
The explicit factor of $E_b - \epsilon_{k_d} - \epsilon_{k_e}$ is included so that $\Psi_k$ is the equal-time wavefunction, i.e., $\Psi_k = \psi_k(t = 0)$.
This ansatz allows us to perform the $\epsilon$ integral straightforwardly, giving a closed equation for $E_b$ and $\Psi_k$:
\begin{equation} \label{eq:simpler_Bethe_Salpeter_equation}
\big( E_b - \epsilon_{k_d} - \epsilon_{k_e} \big) \Psi_k = -\frac{(1 + 2m) t^2}{2N} \sum_q \left( \frac{M_{k_d - q, q} M_{k_e, q}}{E_b - \epsilon_{k_d - q} - \omega_q - \epsilon_{k_e}} + \frac{M_{k_d, -q} M_{k_e + q, -q}}{E_b - \epsilon_{k_d} - \epsilon_{k_e + q} - \omega_q} \right) \Psi_{k-q}.
\end{equation}
\end{widetext}
Eq.~\eqref{eq:simpler_Bethe_Salpeter_equation} is the two-particle Schrodinger equation albeit with an energy-dependent potential.
We find the values of $E_b$ at which it has a non-zero solution, and record the corresponding eigenvector.

Strictly speaking, Eq.~\eqref{eq:Bethe_Salpeter_frequency_ansatz} is not a valid ansatz for $\psi_k(\epsilon)$, i.e., it does not solve the frequency-dependent Eq.~\eqref{eq:Bethe_Salpeter_equation}.
However, it has a clear physical interpretation.
The Fourier transform $\psi_k(t)$ gives the wavefunction for inserting the doublon and holon separated by time $t$ (see Eq.~\eqref{eq:generic_bound_state_wavefunction}).
The poles coming from the single-particle propagators in Eq.~\eqref{eq:Bethe_Salpeter_frequency_ansatz} correspond to the phase factor acquired by the remaining particle during that interval, and our ansatz amounts to neglecting any other time dependence.
This approximation has been applied previously to study holon-holon binding~\cite{Belinicher1997Two}, and we expect it to be qualitatively accurate for our purposes.

Eq.~\eqref{eq:simpler_Bethe_Salpeter_equation} and those preceding it differ from the equations for holon-holon binding in two respects.
First, the holon-holon equations must include exchange terms not found here.
Second, due to the relative phase between the doublon-spinon and holon-spinon vertices, the effective potential in Eq.~\eqref{eq:simpler_Bethe_Salpeter_equation} would have the opposite sign for the holon-holon problem.

\section{Results} \label{sec:results}

\subsection{Single-particle properties}

\begin{figure}[t]
\centering
\includegraphics[width=1.0\columnwidth]{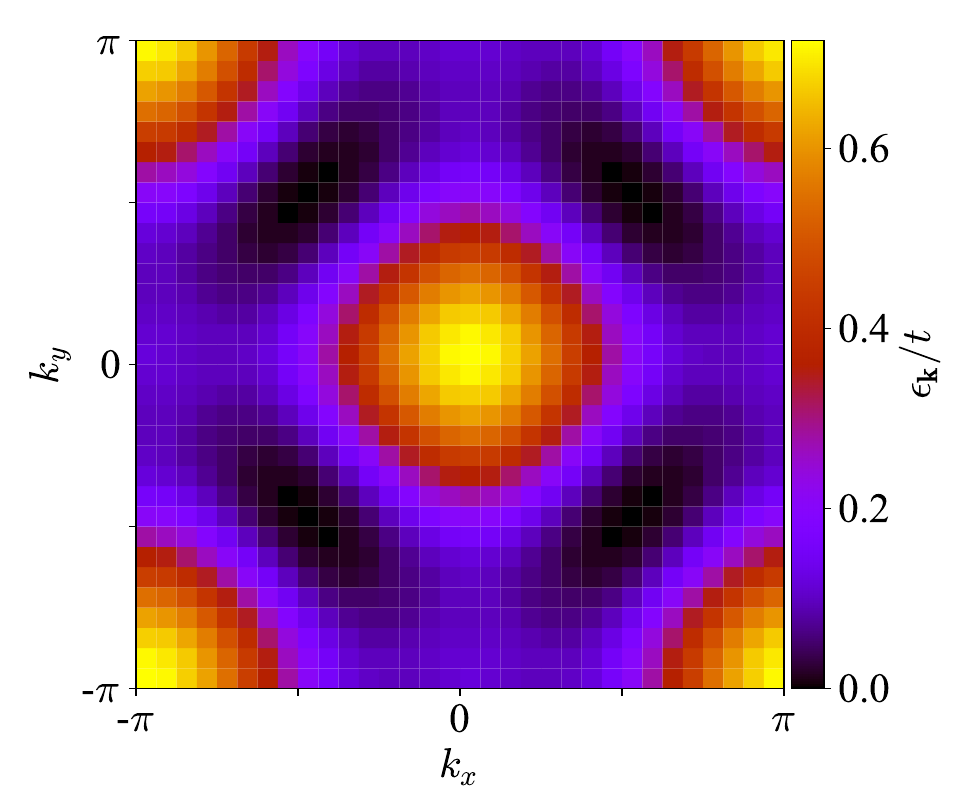}
\caption{Dispersion of individual quasiparticles (both doublons and holons) within the SCBA, for a lattice of size $32 \times 32$. Data shown is for $t/J=2$, $m=0.3$.}
\label{fig:chargon_dispersion}
\end{figure}

We first review the behavior of individual quasiparticles, determined within the SCBA as described above.
Although these calculations have been reported previously, e.g., in Refs.~\cite{SchmittRink1988Spectral,Martinez1991Spin}, it will be useful to reproduce them here.

\begin{figure}[t]
\centering
\includegraphics[width=1.0\columnwidth]{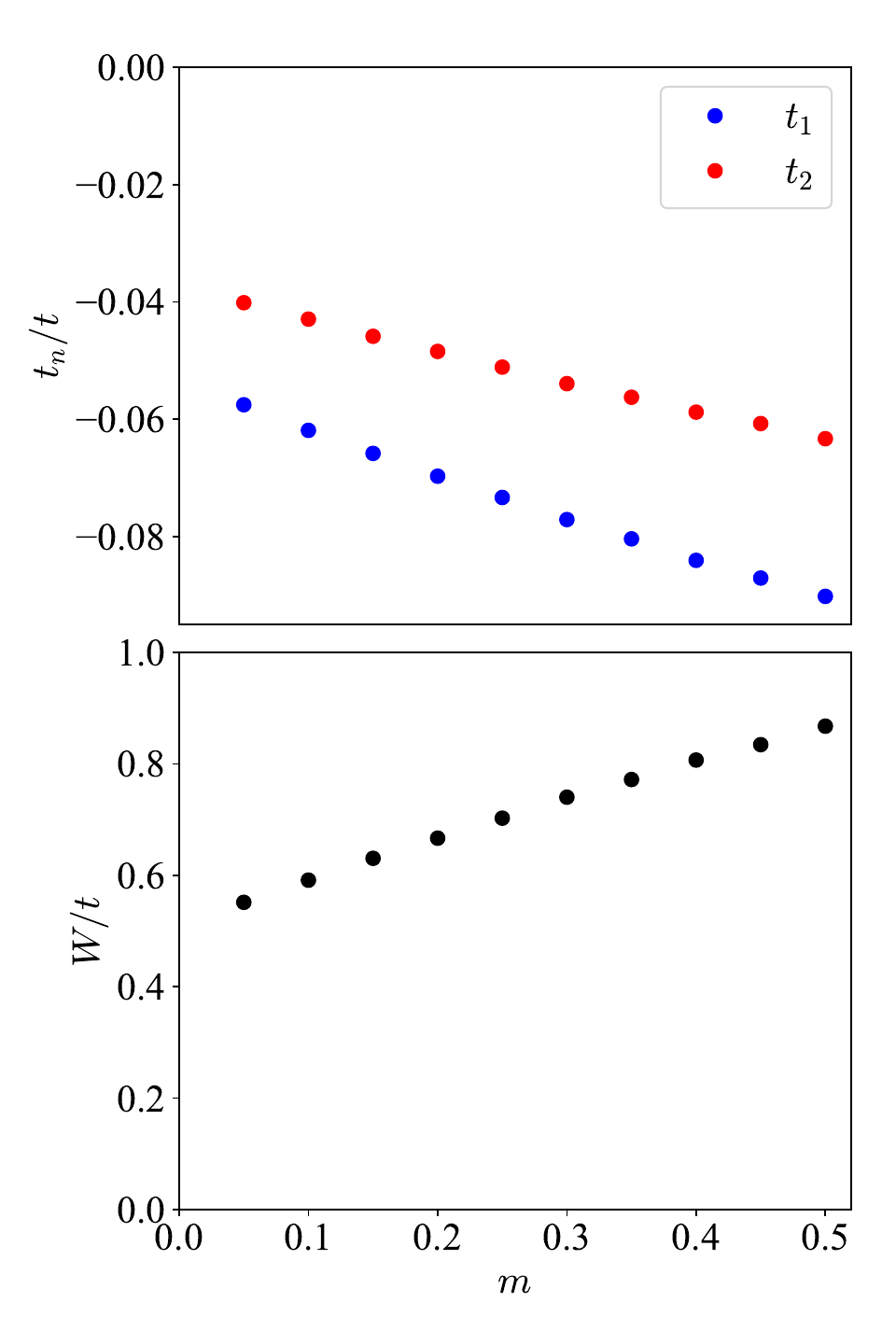}
\caption{(Top) Fitted parameters of the quasiparticle dispersion (Eq.~\eqref{eq:approximate_single_particle_dispersion}) as a function of magnetization, for a lattice of size $32 \times 32$ with $t/J=2$. (Bottom) Quasiparticle bandwidth versus magnetization for the same system.}
\label{fig:chargon_hopping}
\end{figure}

Fig.~\ref{fig:chargon_dispersion} shows the quasiparticle dispersion throughout the Brillouin zone, with the magnetization set to the equilibrium value for concreteness.
As noted above, it can be well-approximated by a next-nearest-neighbor hopping model with amplitude $t_1$ for moving along the diagonals of the lattice and amplitude $t_2$ for moving along the principal axes (see Fig.~\ref{fig:effective_hopping}).
The form of the dispersion is not sensitive to the value of magnetization.
However, the effective hopping amplitudes, which we determine empirically by fitting the computed spectrum to Eq.~\eqref{eq:approximate_single_particle_dispersion}, do depend on $m$ as shown in Fig.~\ref{fig:chargon_hopping}.

Some features of the dispersion can be explained by a simple Hartree-Fock approximation to the original Hamiltonian, in which the Hubbard interaction is replaced by $n_{i \uparrow} \langle n_{i \downarrow} \rangle + \langle n_{i \uparrow} \rangle n_{i \downarrow}$.
Assuming N\'eel order for $\langle n_{i \sigma} \rangle$, the Hamiltonian becomes a tight-binding model on a bipartite lattice with dispersion
\begin{equation} \label{eq:Hartree_Fock_dispersion}
\begin{aligned}
\epsilon_k^{\textrm{(HF)}} =& \, \sqrt{U^2 + 4t^2 \big( \cos{k_x} + \cos{k_y} \big) ^2} \\
\sim & \; U + \frac{2t^2}{U} \big( \cos{k_x} + \cos{k_y} \big) ^2,
\end{aligned}
\end{equation}
using that $t \ll U$.
Up to a constant shift, the second line is equivalent to Eq.~\eqref{eq:approximate_single_particle_dispersion} for the special case $t_1 = 2t_2$.
Note in particular that $t_1, t_2 < 0$.
Thus Hartree-Fock correctly predicts that the band minimum is within the lines $k_k + k_y = \pm \pi$.
It also correctly suggests that the bandwidth should be significantly reduced to $O(J)$ instead of $O(t)$.
However, it \textit{incorrectly} claims that the dispersion is degenerate along the entire magnetic Brillouin zone boundary.
The more sophisticated SCBA resolves this degeneracy, identifying four minima at $(k_x, k_y) = (\pm \pi/2, \pm \pi/2)$.

Recent work on magnetic polarons in the t-J model~\cite{Grusdt2019Microscopic,Bohrdt2020Parton} has made clear that this behavior can be understood by the charge excitations forming spinon-charge bound states (the ``polarons''), held together by strings of displaced spins (much as we have sketched in Fig.~\ref{fig:exciton_cartoon} but with individual charges).
The charges are forced to move on the timescale set by their slower spinon partners, namely $O(1/J)$, and are subject to the bipartite lattice felt by the spinons.
The dispersion results shown above confirm this string picture quite nicely if one interprets them as being for the \textit{polaron} as a whole.
With that in mind, it is rather striking that these three approaches --- SCBA, Hartree-Fock, and the string picture --- all lead to consistent conclusions.

Returning to Fig.~\ref{fig:chargon_hopping}, we see that the bandwidth $W$ increases noticeably as the magnetization increases.
Equivalently, the single-particle mass decreases.
Within the framework of our calculation, the explanation is clear: a doublon/holon can move only if a spinon takes its place (see Eq.~\eqref{eq:t_J_model_def}), and since one spinon factor is always in the direction of the N\'eel magnetization, the doublon/holon hopping term is proportional to $\sqrt{(1 + 2m)/2}$.

\subsection{Exciton properties}

\begin{figure}[t]
\centering
\includegraphics[width=1.0\columnwidth]{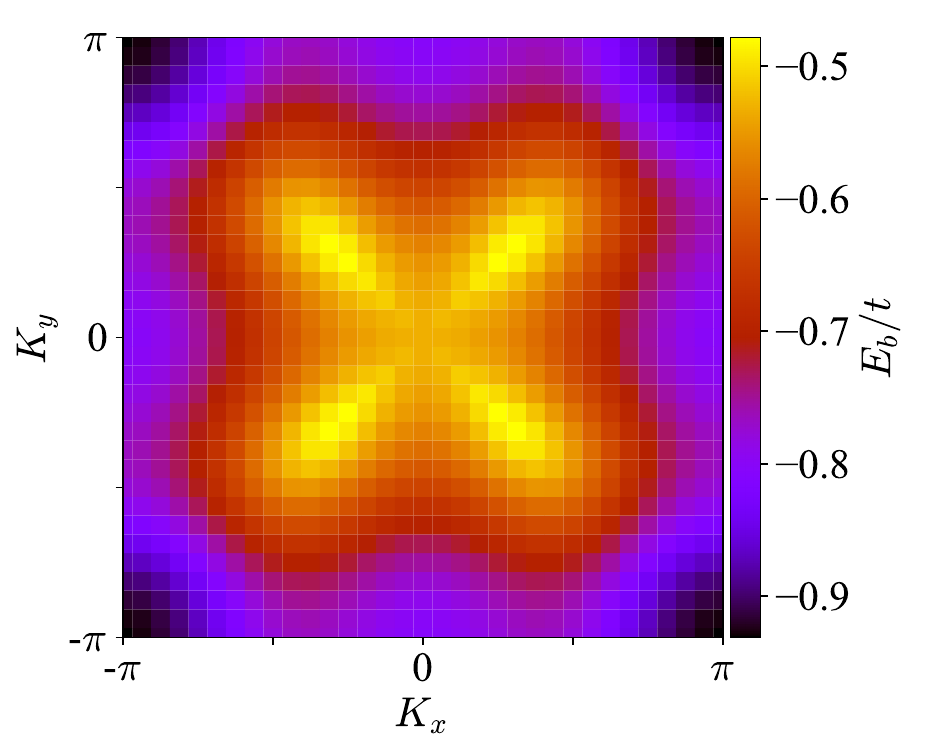}
\caption{Dispersion of the exciton center-of-mass motion, for a lattice of size $32 \times 32$. Data shown is for $t/J=2$, $m=0.3$.}
\label{fig:exciton_dispersion}
\end{figure}

\begin{figure}[t]
\centering
\includegraphics[width=1.0\columnwidth]{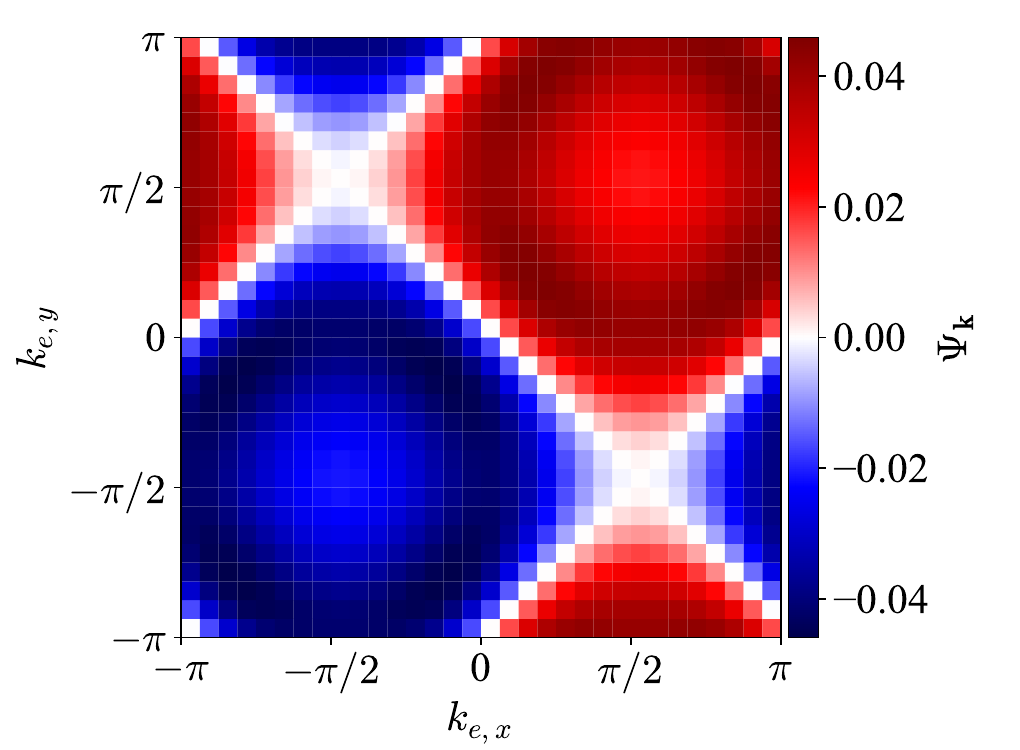}
\caption{Exciton wavefunction in (relative) momentum space at total momentum $(K_x,K_y)=(\pi,\pi)$, for the same system as in Fig.~\ref{fig:exciton_dispersion}.}
\label{fig:exciton_wavefunction}
\end{figure}

\begin{figure}[t]
\centering
\includegraphics[width=1.0\columnwidth]{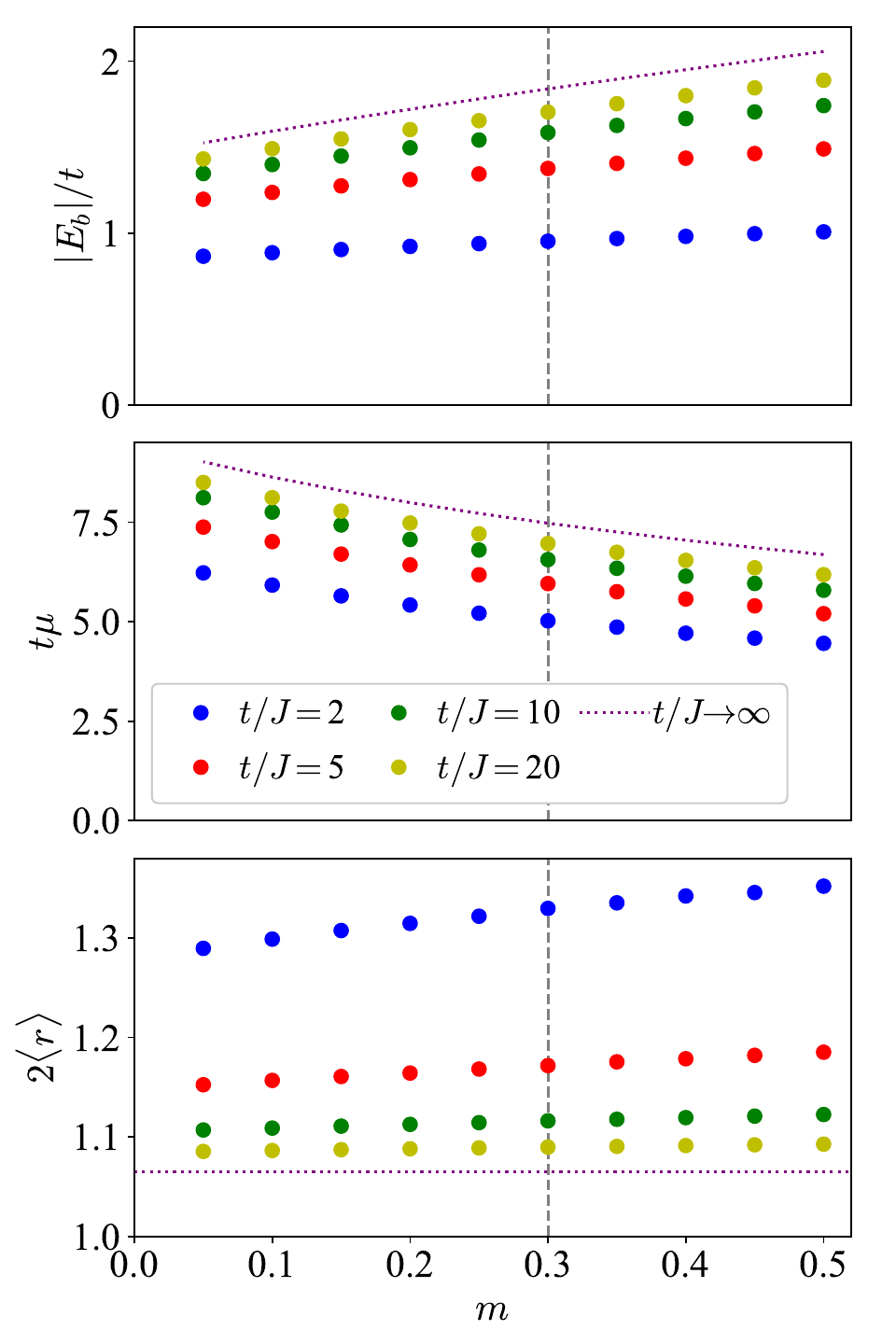}
\caption{Exciton properties as functions of magnetization $m$, for various $t/J$ and a lattice of size $32 \times 32$. The $t/J \rightarrow \infty$ curves are obtained from Eq.~\eqref{eq:large_t_Bethe_Salpeter_equation}. Vertical dashed lines indicate the equilibrium value of $m$. (Top) Binding energy. (Center) Mass. (Bottom) Diameter.}
\label{fig:exciton_properties}
\end{figure}

\begin{figure}[t]
\centering
\includegraphics[width=1.0\columnwidth]{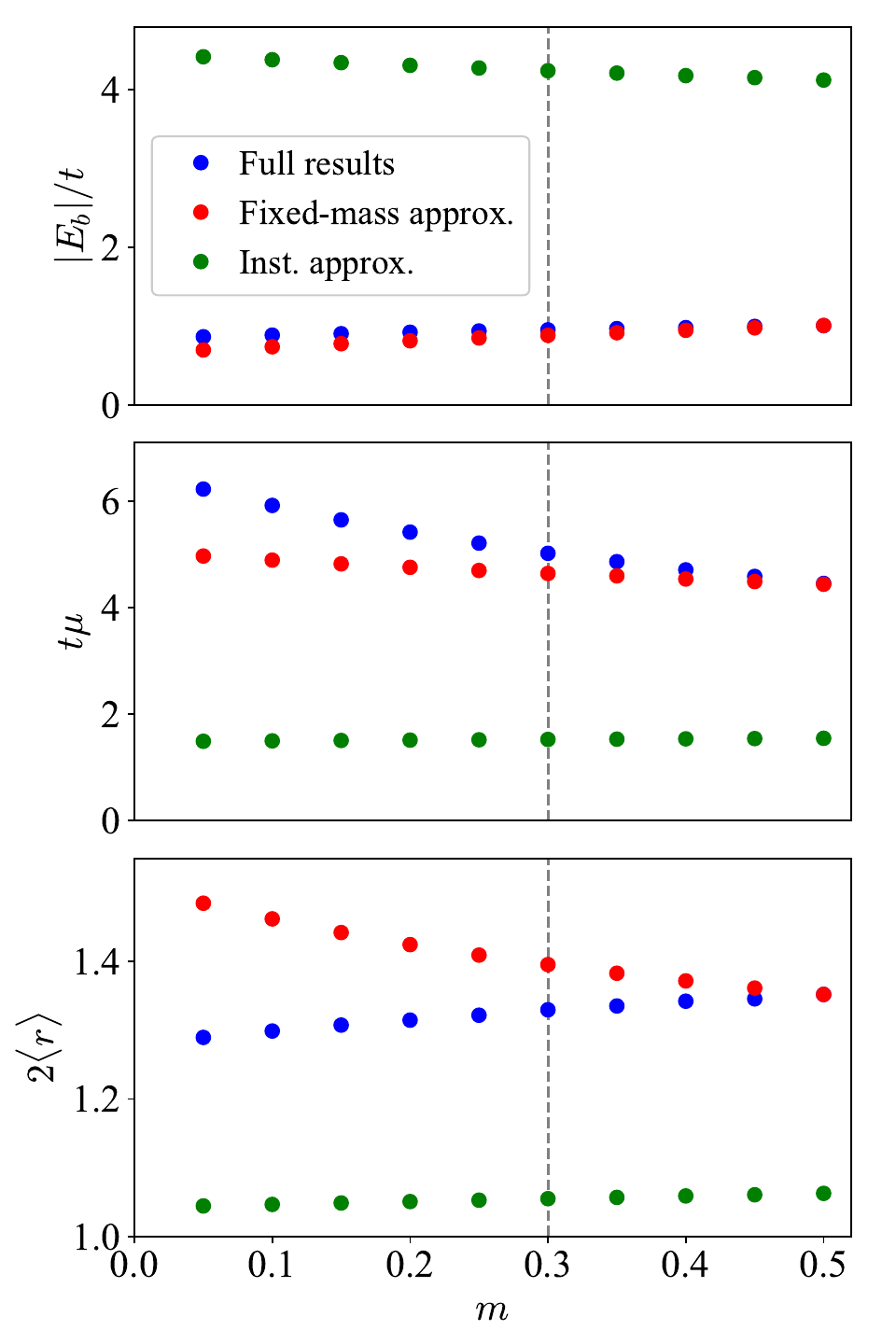}
\caption{Comparison between the full results and the two approximations considered in the text. Data shown is for $t/J = 2$ and a lattice of size $32 \times 32$. Vertical dashed lines indicate the equilibrium value of $m$. For the fixed-mass points, the values of $t_1$ and $t_2$ are set to their values at $m=0.5$ (see Fig.~\ref{fig:chargon_hopping}).}
\label{fig:exciton_comparison}
\end{figure}

We now turn to the exciton properties as functions of magnetization, using the Bethe-Salpeter equation.
All of the quantities presented here are straightforward to compute from the energy $E_b$ and wavefunction $\psi_k$ given by Eq.~\eqref{eq:Bethe_Salpeter_equation}.

Fig.~\ref{fig:exciton_dispersion} shows the energy of the lowest internal state as a function of the center-of-mass momentum $K$.
As was the case for the single-particle dispersion, the shape of the exciton dispersion is not particularly sensitive to the magnetization.
Note that the bottom of the band is not at the origin but rather at $(K_x, K_y) = (\pi, \pi)$.
The wavefunction of the $(\pi, \pi)$ state is shown in Fig.~\ref{fig:exciton_wavefunction}.
We see that it has $p$-wave symmetry, unlike what one would expect for a two-holon bound state.

The binding energy, mass, and radius of the exciton are plotted versus magnetization in Fig.~\ref{fig:exciton_properties}.
We see that as one increases the magnetization $m$, the mass decreases while the binding energy and size increase.
It is interesting to compare these trends with what one would expect for a conventional exciton formed via Coulomb attraction.
In that situation, a decrease in mass is associated with an increase in radius and a decrease in binding energy.
Here, we find a similar relationship between radius and mass, but the binding energy instead scales \textit{inversely} with mass.

Eq.~\eqref{eq:simpler_Bethe_Salpeter_equation} can be simplified further in the large-$t/J$ limit.
We will see that $E_b$ scales as $t$, whereas $\epsilon_k$ and $\omega_q$ are asymptotically smaller~\cite{Martinez1991Spin}.
Thus we can neglect the single-particle and spinon dispersions, leaving the equation
\begin{equation} \label{eq:large_t_Bethe_Salpeter_equation}
\begin{aligned}
E_b^2 \Psi_k =& -\frac{(1 + 2m) t^2}{2N} \sum_q \big( M_{k_d - q, q} M_{k_e, q} \\
& \qquad \qquad \qquad \qquad + M_{k_d, -q} M_{k_e + q, -q} \big) \Psi_{k-q}.
\end{aligned}
\end{equation}
Although still not of the Schrodinger form, Eq.~\eqref{eq:large_t_Bethe_Salpeter_equation} is much simpler to solve than Eq.~\eqref{eq:simpler_Bethe_Salpeter_equation}: the kernel on the right-hand side no longer depends self-consistently on the energy (and as claimed, $E_b \sim t$).
The results obtained from the large-$t/J$ equation are plotted alongside the others in Fig.~\ref{fig:exciton_properties}.

As is clear from Eq.~\eqref{eq:simpler_Bethe_Salpeter_equation}, the spinon-mediated interaction between charges is not instantaneous.
To assess the importance of this retardation, we have compared the results in Fig.~\ref{fig:exciton_properties} to what would be obtained through the static approximation (setting $\omega = 0$ in the kernel of Eq.~\eqref{eq:Bethe_Salpeter_equation}).
The static approximation would predict significantly different results, as seen in Fig.~\ref{fig:exciton_comparison}: the binding energy would instead decrease slightly with magnetization and the mass would increase slightly.
Thus the retardation of the effective interaction is an essential ingredient to the behavior seen here.

Similarly, one can ask whether the trends observed in Fig.~\ref{fig:exciton_properties} are due primarily to changes in the spinon behavior or rather due to the \textit{single}-particle mass, which itself decreases with magnetization.
We have repeated the above calculations under ``fixed-mass'' conditions, in which the single-particle parameters $t_1$ and $t_2$ are kept fixed (to their values at $m = 0.5$) as we vary the magnetization.
Fig.~\ref{fig:exciton_comparison} shows that each of the three observables responds differently.
The binding energy becomes more sensitive to magnetization, indicating that the quasiparticle and spinon properties play antagonistic roles.
On the other hand, the exciton mass becomes less sensitive -- the change to the effective interaction suppresses the mass by itself.
Finally, the exciton radius shows the reverse behavior to before, instead decreasing with magnetization (although the size remains quite small in absolute terms).

The recent pump-probe experiments in Refs.~\cite{Alpichshev2015Confinement,Alpichshev2017Origin} have investigated how excitons are influenced by magnetic order in the Mott insulator Na\textsubscript{2}IrO\textsubscript{3}.
Our results support their interpretation in some aspects but not in others.
In Ref.~\cite{Alpichshev2015Confinement}, the authors observe an increase in the fraction of bound excitations when below the N\'eel temperature, which they attribute to an increase in the exciton binding energy.
Fig.~\ref{fig:exciton_properties} shows that magnetic order does indeed increase the binding energy.
On the other hand, Ref.~\cite{Alpichshev2017Origin} demonstrates that the relaxational dynamics following a pump are slower in the presence of magnetic order.
This is attributed to the mass increasing with magnetization, yet we have observed the opposite (consistent with past works calculating the dependence on $J/t$~\cite{Kane1989Motion,VonSzczepanski1990Single,Martinez1991Spin}).
Given the highly non-equilibrium nature of the experiments, as well as the approximations inherent in an analytical approach, further investigation is clearly needed.

 Finally, let us compare the present calculation of doublon-holon binding to that of holon-holon binding, which is obviously of significant interest in its own right~\cite{Lee2006Doping,Plakida2010}.
Clearly the two have much in common, yet there are two important differences.
First, the integral equation which determines the two-particle Green's function (Fig.~\ref{fig:Bethe_Salpeter}) has an additional exchange term due to the indistinguishability of the holons.
Second, even the direct term comes with an extra minus sign, i.e., the effective interaction is of opposite sign.
The sign can be removed by redefining the hole operator on one sublattice, but the additional phase may modify further results depending on the application.
It is important to keep these distinctions in mind when relating the present results to the high-$T_c$ literature.

\section{Conclusion} \label{sec:conclusion}

We have studied the role that magnetic order plays in the formation of excitons within Mott insulators, using the Hubbard model as a concrete Hamiltonian.
The binding energy increases in the presence of (antiferromagnetic) magnetization, whereas the exciton mass decreases.
The size of the exciton increases slightly, yet the radius is never more than a lattice spacing. Using the standard classification, these are Frenkel excitons regardless of magnetic order.

In addition, we have established that the trends observed here require a detailed understanding of the many-body dynamics in these systems. Retardation effects in the effective spinon-mediated interaction are essential. Furthermore, the constituent charge and spin excitations are each affected separately by the background magnetic order, in ways cooperative for some exciton properties but antagonistic for others.

It must be noted that despite the complexity, there are significant limitations to our approach.
In particular, we have made approximations in the spirit of linear spin-wave theory, which is only justified at large spin $S$ and full magnetization (neither of which we assume here).
Thus we do not expect these results to be quantitatively accurate --- we instead view this analysis as expressing our physical intuition regarding Mott excitons in the language of slave particles, from which we can make sharp predictions to be verified or falsified by more systematic investigations.

As an outlook, the predictions made here will be important when analyzing recent and future experiments on the optical properties of strongly correlated electronic materials.
The existing experiments are quite complex, and require interpretations of their own.
Our results agree with those interpretations in some respects but disagree in others.
A complete understanding of the systems will require numerous approaches, both experimental and theoretical, including but not limited to the one described here.

Particularly promising are the recent experiments on fermionic atoms in optical lattices~\cite{Hilker2017Revealing,Salomon2019Direct,Vijayan2020Time}.
Since ultracold gases do not have many of the complicating features found in condensed matter systems, we expect that this will be a valuable direction to explore further.
It is also likely that our conclusions, being based on the single-band nearest-neighbor Hubbard model, are more applicable to those systems than to materials such as the iridates. Importantly, current quantum gas microscopes allow one to directly create localized doublons and holons via optical tweezers, and  reliably measure the spin correlation functions~\cite{Koepsell2019Imaging}. Such an unprecedented direct access to the system microscopics will provide a powerful way of investigating many-body excitons.

\section{Acknowledgements}
We would like to thank Eugene Demler, Immanuel Bloch, Andrew Allocca, Zachary Raines, Yang-Zhi Chou, Fabian Grudst, and Annabelle Bohrdt for stimulating discussions and suggestions. This work was supported by the NSF Physics Frontier Center at the Joint Quantum Institute, the NSF DMR-1613029 and US-ARO contracts W911NF1310172 (T.-S.H), W911NF2010232,  NRC Research Associateship award at the National Institute of Standards and Technology (CLB), AFOSR-MURI FA9550-16-1-0323 (M.H.), DOE-BES award DESC0001911 (V.G.) and the Simons Foundation. M.H. \& V.G. acknowledge the hospitality of KITP-UCSB, which is supported in part by Grant No. NSF PHY-1748958.

\bibliography{Hubbard_Exciton_v2_Biblio}

\end{document}